\documentclass[final,3p,times]{elsarticle}

\usepackage{amssymb}
 \usepackage{amsthm}
\usepackage{cancel}

\usepackage{hyperref}
\usepackage{verbatim}

\usepackage{mathrsfs}
\usepackage{amsmath}
\usepackage{bm}
\usepackage{graphicx}
\usepackage{color}

 \biboptions{square,sort&compress,comma}

\newcounter{bla}

\journal{Computer Physics Communications}

\begin{document}

\begin{frontmatter}



\title{BO 2.0: Plasma Wave and Instability Analysis with Enhanced Polarization Calculations}

\date{\today}


\author[label1,label2]{Hua-sheng XIE}
\ead{huashengxie@gmail.com, xiehuasheng@enn.cn}

\author[label3]{Richard Denton}
\ead{redenton@gmail.com} 

\author[label4,label5]{Jin-song Zhao}
\ead{js\_zhao@pmo.ac.cn}

\author[label4,label5]{Wen Liu}

\address[label1]{Hebei Key Laboratory of Compact Fusion, Langfang 065001, China}
\address[label2]{ENN Science and Technology Development Co., Ltd., Langfang 065001, China}
\address[label3]{Department of Physics and Astronomy, Dartmouth College, Hanover, New Hampshire, USA}
\address[label4]{Key Laboratory of Planetary Sciences, Purple Mountain Observatory, Chinese Academy of Sciences, Nanjing 210008, People`s Republic of China}
\address[label5]{School of Astronomy and Space Science, University of Science and Technology of China, Hefei 230026, People`s Republic of China}

\begin{abstract}
Besides the relation between the wave vector $\bm k$ and the complex frequency $\omega$, wave polarization is useful for characterizing the properties of a plasma wave.  The polarization of the electromagnetic fields, $\delta \bm E$ and $\delta \bm B$, have been widely used in plasma physics research. Here, we derive equations for the density and velocity perturbations, $\delta n_s$ and $\delta{\bm v}_s$, respectively, of each species in the electromagnetic kinetic plasma dispersion relation by using their relation to the species current density perturbation $\delta {\bm J}_s$. Then we compare results with those of another commonly used plasma dispersion code (WHAMP) and with those of a multi-fluid plasma dispersion relation. We also summarize a number of useful polarization quantities, such as magnetic ellipticity, orientation of the major axis of the magnetic ellipse, various ratios of field energies and kinetic energies, species compressibility, parallel phase ratio, Alfv\'en-ratio, etc., which are useful for plasma physics research, especially for space plasma studies. This work represents an extension of the BO electromagnetic dispersion code [H.S. Xie, Comput. Phys. Comm. 244 (2019) 343-371] to enhance its calculation of polarization and to include the capability of solving the electromagnetic magnetized multi-fluid plasma dispersion relation.

\end{abstract}

\end{frontmatter}



\noindent
{\bf PROGRAM SUMMARY}

\begin{small}
\noindent
{\em Program Title:}  BO 2.0                                       \\
{\em Licensing provisions:}        BSD 3-clause                            \\
{\em Programming language:}    Matlab                               \\
{\em Journal reference of previous version:} [1] H.S. Xie, BO: A unified tool for plasma waves and instabilities analysis, Comput. Phys. Comm. 244 (2019) 343-371. [2]  H.S. Xie, Y. Xiao, PDRK: A General Kinetic Dispersion Relation Solver for Magnetized Plasma, Plasma Sci. Technol. 18 (2) (2016) 97. [3] H. S. Xie, PDRF: A general dispersion relation solver for magnetized multi-fluid plasma, Comput. Phys. Comm. 185 (2014) 670-675.                 \\
{\em Does the new version supersede the previous version?:}  Yes \\
{\em Reasons for the new version:} Enhance the code capability, especially to support the calculation of density and velocity perturbations. Also, the multi-fluid and kinetic versions are combined into one version.\\
{\em Summary of revisions:}* In this new version, multi-fluid model is included as one option. The density and velocity perturbations of kinetic versions are also supported. Many useful polarizations are included.\\
{\em Nature of problem:} The linear fluid and kinetic waves and instabilities in plasma can be described by dispersion relations. The challenges are to provide a dispersion relation as general as possible and to obtain all the solutions of it, which is the goal of BO. The BO code provides a unified numerically solvable framework for kinetic and multi-fluid plasma dispersion relations, which greatly extends the standard ones, with an arbitrary number of species.\\
{\em Solution method: }  Transforming the dispersion relation to an equivalent matrix eigenvalue problem and find all the solutions using standard matrix eigenvalue library function.\\
{\em Additional comments including Restrictions and Unusual features (approx. 50-250 words):} Kinetic relativistic effects are not included in the present version yet.\\

\end{small}


\section{Introduction}

Plasma is a combination of particles and electromagnetic fields. The electromagnetic fields are usually described by the Maxwell equations, whereas particles can be described by either a kinetic model using the distribution function, $f_s({\bm v,t})$, or a fluid model that employs velocity moments of the distribution function. It is well known that, for a uniform plasma, the linear plasma dispersion relation can be solved using
\begin{eqnarray}\label{eq:DE}
{\bm D}\cdot \delta {\bm E}=0,
\end{eqnarray}
with
\begin{eqnarray}\label{eq:DR}
|D(\omega,{\bm k})|=0,
\end{eqnarray}
where ${\bm D}$ is a 3-by-3 matrix tensor and $\delta{\bm E}=(\delta E_x, \delta E_y, \delta E_z)$ is the perturbed electric field. For a given wave vector ${\bm k}$, we solve the dispersion relation Eq.(\ref{eq:DR}), which yields the complex frequency, $\omega=\omega_r+i\omega_i$. Then, we obtain the matrix elements of ${\bm D}$ from $\omega$ and solve Eq. (\ref{eq:DE}) for the perturbed electric field $\delta {\bm E}$. We can then calculate the perturbed magnetic field $\delta{\bm B}$ and current density $\delta{\bm J}$ using the Maxwell equations. 



Besides the perturbed electromagnetic fields, the plasma waves can carry the perturbed density $\delta n_s$ and velocity $\delta {\bm v}_s$, which are widely used for the wave mode identification. Here ``s" denotes the particle species.
$\delta n_s$ and $\delta {\bm v}_s$ were not given in the kinetic dispersion code BO v1.0 \cite{Xie2019,Xie2016}, which has been shown to be a powerful tool for studying plasma waves and instabilities in the solar-terrestrial plasmas \cite{Sun2019,Sun2020}.
The major purpose of this work is to derive expressions for $\delta n_s$ and $\delta{\bm v}_s$ using the kinetic dispersion relation, and check the validity of the approach by comparing results with those of another commonly used electromagnetic dispersion code (WHAMP \cite{Ronnmark1982}) and with those of a multi-fluid plasma model \cite{Xie2014}. In section \ref{sec:eqn}, we derive the equations and show how we implement them in the BO kinetic dispersion code \cite{Xie2019,Xie2016} and the PDRK fluid dispersion code \cite{Xie2014}. In section \ref{sec:bench}, we benchmark the results using two independent kinetic solvers and with the multi-fluid solver PDRF. In section \ref{sec:summ}, we give a summary with some discussion. In the Appendices, we list useful polarization quantities calculated in BO and give a summary of the updated model used in PDRF. All of these updates are summarized to the new version BO v2.0 (https://github.com/hsxie/bo).

\section{How to calculate the perturbed density, velocity and plasma current in BO}\label{sec:eqn}


\subsection{Perturbed density and velocity}

In the plasma kinetic model, the density and velocity are zeroth and first order moment of the velocity distribution function, respectively. The perturbed density is given by $\delta n_s=n_{s0} \int \delta f_s dv^3$, where $n_{s0}$ is the zeroth order density and $\delta f_s$ is the first order perturbed velocity distribution function. The controlling equation for the perturbed density can be obtained through performing zeroth order moment for the linear Vlasov equation, 
\begin{equation}
\partial_t\delta n_s+\nabla\cdot(\delta {\bm J}_s/q_s)=0,
\end{equation}
where $\delta {\bm J}_s=n_{s0} q_s \int {\bm v}\delta f_s dv^3$ is the perturbed plasma current. The perturbed plasma current can be expressed in terms of the perturbed fluid density and velocity,
\begin{equation}
\delta{\bm J}_s=q_sn_{s0}\delta{\bm v}_s+q_s\delta n_s{\bm v}_{ds},
\end{equation}
or
\begin{eqnarray}
\left\{
\begin{array}{lcl}
\delta J_{sx}&=&q_sn_{s0}\delta v_{sx}+q_s\delta n_sv_{dsx},\\
\delta J_{sy}&=&q_sn_{s0}\delta v_{sy}+q_s\delta n_sv_{dsy},\\
\delta J_{sz}&=&q_sn_{s0}\delta v_{sz}+q_s\delta n_sv_{dsz},
\end{array}\right.
\end{eqnarray}
where ${\bm v}_{ds}=(v_{dsx},v_{dsy},v_{dsz})$ denotes the zeroth order drift velocity, and directions of $x$ and $y$ axes are perpendicular to the background magnetic field ${\bm B_0}=B_0{\bm z}$. It should be noted that one of advantages of BO \cite{Xie2019} and PDRF \cite{Xie2014} is that the three components of ${\bm v}_{ds}$ are included in these two solvers. 


Using Eqs. (3) and (5), we can directly obtain $\delta n_s$ and $\delta {\bm v}_s$ once the dispersion relation of one plasma wave mode and $\delta{\bm J}_s$ are known. Under the plane wave assumption, i.e., $\partial_t\to-i\omega$, $\nabla\to i{\bm k}$, one readily obtains
\begin{eqnarray}
\omega\delta n_s=\frac{1}{q_s}{\bm k}\cdot\delta{\bm J}_s.
\end{eqnarray}
Since the wavevector is given as ${\bm k}=(k_x,0,k_z)$ in BO \cite{Xie2019}, the controlling equations for the perturbed density and velocity are
\begin{eqnarray}\label{eq:calnv}
\left\{
\begin{array}{lcl}
\delta n_s&=&\frac{1}{\omega q_s}(k_x\delta J_{sx}+k_z\delta J_{sz}),\\
\delta v_{sx}&=&\frac{1}{q_sn_{s0}}\Big[\delta J_{sx}-\frac{1}{\omega}(k_x\delta J_{sx}+k_z\delta J_{sz})v_{dsx}\Big],\\
\delta v_{sy}&=&\frac{1}{q_sn_{s0}}\Big[\delta J_{sy}-\frac{1}{\omega}(k_x\delta J_{sx}+k_z\delta J_{sz})v_{dsy}\Big],\\
\delta v_{sz}&=&\frac{1}{q_sn_{s0}}\Big[\delta J_{sz}-\frac{1}{\omega}(k_x\delta J_{sx}+k_z\delta J_{sz})v_{dsz}\Big].
\end{array}\right.
\end{eqnarray}
Here we note that $\delta{\bm v}_s=\int dv^3{\bm v}\delta f_s = \delta {\bm J}_s/n_{s0}/q_s$ in motionless plasmas where ${\bm v}_{ds}=0$, and $\delta{\bm v}_s\neq \int dv^3{\bm v}\delta f_s$ in a plasma where ${\bm v}_{ds}\neq 0$.




\subsection{Perturbed plasma current}


In this subsection, we will discuss how to calculate the plasma current $\delta{\bm J}_s$ in BO/PDRK \cite{Xie2019,Xie2016}.
The first approach for giving $\delta{\bm J}_s$ is through $\delta{\bm J}_s={\bm \sigma}_s\cdot \delta {\bm E}$, where the conductivity tensor ${\bm \sigma}_s$ in BO/PDRK can be obtained by the following procedures. Eq. (129) in Ref. \cite{Xie2019} gives the relation of the total plasma current and electric field in BO/PDRK

\begin{equation}\label{eq:JmE}
    \left( \begin{array}{c} \delta J_x^m \\  \delta J_y^m \\  \delta J_z^m\end{array}\right)
    =-i\epsilon_0\left( \begin{array}{ccc}
    \frac{b_{11}^m}{\omega}+\sum_{snj}\frac{b_{snj11}}{\omega-c_{snj}} & \frac{b_{12}^m}{\omega}+\sum_{snj}\frac{b_{snj12}}{\omega-c_{snj}}
    & \frac{b_{13}^m}{\omega}+\sum_{snj}\frac{b_{snj13}}{\omega-c_{snj}}\\
    \frac{b_{21}^m}{\omega}+\sum_{snj}\frac{b_{snj21}}{\omega-c_{snj}} & \frac{b_{22}^m}{\omega}+\sum_{snj}\frac{b_{snj22}}{\omega-c_{snj}}
    & \frac{b_{23}^m}{\omega}+\sum_{snj}\frac{b_{snj23}}{\omega-c_{snj}} \\
    \frac{b_{31}^m}{\omega}+\sum_{snj}\frac{b_{snj31}}{\omega-c_{snj}} & \frac{b_{32}^m}{\omega}+\sum_{snj}\frac{b_{snj32}}{\omega-c_{snj}}
    & \frac{b_{33}^m}{\omega}+\sum_{snj}\frac{b_{snj33}}{\omega-c_{snj}}
    \end{array}\right) \left( \begin{array}{c} \delta E_x \\  \delta E_y \\
     \delta E_z\end{array}\right).
\end{equation}
with coefficients
\begin{equation}\label{eq:JEmatm}
    \left\{ \begin{array}{ccc}
   b_{snj11} =  \sum_{\sigma}r_{s\sigma}\omega_{ps}^2p_{11snj}/c_{snj},~
    b_{11}^m = - \sum_{s}^{s=m}\omega_{ps}^2\sum_{\sigma}r_{s\sigma}[\sum_n A_{nbs\sigma}\frac{n\omega_{cs}}{k_x  v_{\perp ts\sigma}^2}( \frac{n\omega_{cs}}{k_x }+v_{dsx})+ \sum_{nj}p_{11snj}/c_{snj}],\\
    
    b_{snj12} =  \sum_{\sigma}r_{s\sigma}\omega_{ps}^2p_{12snj}/c_{snj},~~~~~~
    b_{12}^m = - \sum_{s}^{s=m}\omega_{ps}^2[\sum_{\sigma}r_{s\sigma}\sum_{nj}p_{12snj}/c_{snj}],\\

    b_{snj21} =  \sum_{\sigma}r_{s\sigma}\omega_{ps}^2p_{21snj}/c_{snj},~~~~~~
    b_{21}^m = - \sum_{s}^{s=m}\omega_{ps}^2[\sum_{\sigma}r_{s\sigma}\sum_{nj}p_{21snj}/c_{snj}],\\

   b_{snj22} =  \sum_{\sigma}r_{s\sigma}\omega_{ps}^2p_{22snj}/c_{snj},~~
    b_{22}^m = - \sum_{s}^{s=m}\omega_{ps}^2\sum_{\sigma}r_{s\sigma}[\sum_n(C_{nbs\sigma}+i\frac{v_{dsy}}{v_{\perp ts\sigma} }B_{nbs\sigma})+\sum_{nj}p_{22snj}/c_{snj}],\\

   b_{snj13} =  \sum_{\sigma}r_{s\sigma}\omega_{ps}^2p_{13snj}/c_{snj},~~~~~~
    b_{13}^m = - \sum_{s}^{s=m}\omega_{ps}^2[\sum_{\sigma}r_{s\sigma}\sum_{nj}p_{13snj}/c_{snj}],\\

    b_{snj31} =  \sum_{\sigma}r_{s\sigma}\omega_{ps}^2p_{31snj}/c_{snj},~~~~~~
    b_{31}^m = - \sum_{s}^{s=m}\omega_{ps}^2[\sum_{\sigma}r_{s\sigma}\sum_{nj}p_{31snj}/c_{snj}],\\

   b_{snj23} =  \sum_{\sigma}r_{s\sigma}\omega_{ps}^2p_{23snj}/c_{snj},~~~~~~
    b_{23}^m = - \sum_{s}^{s=m}\omega_{ps}^2[\sum_{\sigma}r_{s\sigma}\sum_{nj}p_{23snj}/c_{snj}],\\

   b_{snj32} =  \sum_{\sigma}r_{s\sigma}\omega_{ps}^2p_{32snj}/c_{snj},~~~~~~
    b_{32}^m = - \sum_{s}^{s=m}\omega_{ps}^2[\sum_{\sigma}r_{s\sigma}\sum_{nj}p_{32snj}/c_{snj}],\\

  b_{snj33} =  \sum_{\sigma}r_{s\sigma}\omega_{ps}^2p_{33snj}/c_{snj},~~~~~~
    b_{33}^m = - \sum_{s}^{s=m}\omega_{ps}^2 \sum_{\sigma}r_{s\sigma}[\sum_n\frac{1}{2}A_{n0\sigma}+\sum_{nj}p_{33snj}/c_{snj}],\\

  c_{snj} = c_{snj}=k_zv_{dsz}+n\omega_{cs}+k_xv_{dsx}-i\nu_s+k_zv_{zts}c_j.
    \end{array}\right.
\end{equation}
The definition for variables in Eqs. (8) and (9) can be found in Ref. \cite{Xie2019}. 

To implement Eq. (\ref{eq:calnv}), we need to separate the contribution from each species of $\delta {\bm J}=\sum_s \delta{\bm J}_s=\sum_s{\bm \sigma}_s\cdot \delta {\bm E}$. To do this, we use a relation of $\frac{b_{11}}{\omega}=\sum_s\frac{b_{11s}}{\omega}$, and then we rewrite Eqs. (\ref{eq:JmE}) and (\ref{eq:JEmatm}) as
\begin{eqnarray}\label{eq:JmEnew}\nonumber
    \left( \begin{array}{c}\delta J_x^m \\ \delta J_y^m \\ \delta J_z^m\end{array}\right)
    &=&-i\epsilon_0\left( \begin{array}{ccc}
   \sum_s \frac{b_{s11}^m}{\omega}+\sum_{snj}\frac{b_{snj11}}{\omega-c_{snj}} & \sum_s\frac{b_{s12}^m}{\omega}+\sum_{snj}\frac{b_{snj12}}{\omega-c_{snj}}
    &\sum_s \frac{b_{s13}^m}{\omega}+\sum_{snj}\frac{b_{snj13}}{\omega-c_{snj}}\\
    \sum_s\frac{b_{s21}^m}{\omega}+\sum_{snj}\frac{b_{snj21}}{\omega-c_{snj}} & \sum_s\frac{b_{s22}^m}{\omega}+\sum_{snj}\frac{b_{snj22}}{\omega-c_{snj}}
    & \sum_s\frac{b_{s23}^m}{\omega}+\sum_{snj}\frac{b_{snj23}}{\omega-c_{snj}} \\
   \sum_s \frac{b_{s31}^m}{\omega}+\sum_{snj}\frac{b_{snj31}}{\omega-c_{snj}} & \sum_s\frac{b_{s32}^m}{\omega}+\sum_{snj}\frac{b_{snj32}}{\omega-c_{snj}}
    & \sum_s\frac{b_{s33}^m}{\omega}+\sum_{snj}\frac{b_{snj33}}{\omega-c_{snj}}
    \end{array}\right) \left( \begin{array}{c}\delta E_x \\ \delta E_y \\
    \delta E_z\end{array}\right)\\
    &=&\sum_s{\bm \sigma}_s^m\cdot \left( \begin{array}{c}\delta E_x \\ \delta E_y \\
    \delta E_z\end{array}\right)
\end{eqnarray}
with the coefficients
\begin{equation}\label{eq:JEmatmnew}
    \left\{ \begin{array}{ccc}
   b_{snj11} =  \sum_{\sigma}r_{s\sigma}\omega_{ps}^2p_{11snj}/c_{snj},~
    b_{s11}^m = - \omega_{ps}^2\sum_{\sigma}r_{s\sigma}[\sum_n A_{nbs\sigma}\frac{n\omega_{cs}}{k_x  v_{\perp ts\sigma}^2}( \frac{n\omega_{cs}}{k_x }+v_{dsx})+ \sum_{nj}p_{11snj}/c_{snj}],\\
    
    b_{snj12} =  \sum_{\sigma}r_{s\sigma}\omega_{ps}^2p_{12snj}/c_{snj},~~~~~~
    b_{s12}^m = - \omega_{ps}^2[\sum_{\sigma}r_{s\sigma}\sum_{nj}p_{12snj}/c_{snj}],\\

    b_{snj21} =  \sum_{\sigma}r_{s\sigma}\omega_{ps}^2p_{21snj}/c_{snj},~~~~~~
    b_{s21}^m = - \omega_{ps}^2[\sum_{\sigma}r_{s\sigma}\sum_{nj}p_{21snj}/c_{snj}],\\

   b_{snj22} =  \sum_{\sigma}r_{s\sigma}\omega_{ps}^2p_{22snj}/c_{snj},~~
    b_{s22}^m = - \omega_{ps}^2\sum_{\sigma}r_{s\sigma}[\sum_n(C_{nbs\sigma}+i\frac{v_{dsy}}{v_{\perp ts\sigma} }B_{nbs\sigma})+\sum_{nj}p_{22snj}/c_{snj}],\\

   b_{snj13} =  \sum_{\sigma}r_{s\sigma}\omega_{ps}^2p_{13snj}/c_{snj},~~~~~~
    b_{s13}^m = - \omega_{ps}^2[\sum_{\sigma}r_{s\sigma}\sum_{nj}p_{13snj}/c_{snj}],\\

    b_{snj31} =  \sum_{\sigma}r_{s\sigma}\omega_{ps}^2p_{31snj}/c_{snj},~~~~~~
    b_{s31}^m = - \omega_{ps}^2[\sum_{\sigma}r_{s\sigma}\sum_{nj}p_{31snj}/c_{snj}],\\

   b_{snj23} =  \sum_{\sigma}r_{s\sigma}\omega_{ps}^2p_{23snj}/c_{snj},~~~~~~
    b_{s23}^m = - \omega_{ps}^2[\sum_{\sigma}r_{s\sigma}\sum_{nj}p_{23snj}/c_{snj}],\\

   b_{snj32} =  \sum_{\sigma}r_{s\sigma}\omega_{ps}^2p_{32snj}/c_{snj},~~~~~~
    b_{s32}^m = - \omega_{ps}^2[\sum_{\sigma}r_{s\sigma}\sum_{nj}p_{32snj}/c_{snj}],\\

  b_{snj33} =  \sum_{\sigma}r_{s\sigma}\omega_{ps}^2p_{33snj}/c_{snj},~~~~~~
    b_{s33}^m = - \omega_{ps}^2 \sum_{\sigma}r_{s\sigma}[\sum_n\frac{1}{2}A_{n0\sigma}+\sum_{nj}p_{33snj}/c_{snj}],\\

  c_{snj} = c_{snj}=k_zv_{dsz}+n\omega_{cs}+k_xv_{dsx}-i\nu_s+k_zv_{zts}c_j.
    \end{array}\right.
\end{equation}
Consequently, we have
\begin{eqnarray}\label{eq:JsmE}
    \delta{\bm J}_s^m={\bm \sigma}_s^m\cdot\delta{\bm E},~~~{\bm \sigma}_s^m=-i\epsilon_0\left( \begin{array}{ccc}
    \frac{b_{s11}^m}{\omega}+\sum_{nj}\frac{b_{snj11}}{\omega-c_{snj}} & \frac{b_{s12}^m}{\omega}+\sum_{nj}\frac{b_{snj12}}{\omega-c_{snj}}
    & \frac{b_{s13}^m}{\omega}+\sum_{nj}\frac{b_{snj13}}{\omega-c_{snj}}\\
    \frac{b_{s21}^m}{\omega}+\sum_{nj}\frac{b_{snj21}}{\omega-c_{snj}} & \frac{b_{s22}^m}{\omega}+\sum_{nj}\frac{b_{snj22}}{\omega-c_{snj}}
    & \frac{b_{s23}^m}{\omega}+\sum_{nj}\frac{b_{snj23}}{\omega-c_{snj}} \\
    \frac{b_{s31}^m}{\omega}+\sum_{nj}\frac{b_{snj31}}{\omega-c_{snj}} & \frac{b_{s32}^m}{\omega}+\sum_{nj}\frac{b_{snj32}}{\omega-c_{snj}}
    & \frac{b_{s33}^m}{\omega}+\sum_{nj}\frac{b_{snj33}}{\omega-c_{snj}}
    \end{array}\right).
\end{eqnarray}
We can use Eq. (\ref{eq:JsmE}) to obtain $\delta{\bm J}_s$. Since $\delta{\bm E}$ is known, the first approach requires solving the above 3-by-3 tensor for each species.




The second approach would be more convenient for obtaining $\delta{\bm J}_s$ based on the fact that a matrix eigenvalue method is used in BO/PDRK. For example, as given in Eq. (132) of Ref. \cite{Xie2019}, the perturbed current in $x$ direction is
\begin{equation}\label{eq:Jjvv}
i\delta J_x\epsilon_0=j_x+\sum_{snj}^{s=m}v_{snjx}+\sum_{sj\sigma}^{s=u}v_{sj\sigma x},
\end{equation}
where $j_x$, $v_{snjx}$ and $v_{sj\sigma x}$ have been solved along with $\delta{\bm E}$ and $\delta{\bm B}$. $\delta J_{sx}$ can be directly obtained once $j_x$, $v_{snjx}$ and $v_{sj\sigma x}$ for each species $s$ are known. Similarly, we can obtain $\delta J_{sy}$ and $\delta J_{sz}$.

The quantities $v_{snjx}$ and $v_{sj\sigma x}$ are species quantities, but $j_x$ was not in the original version of BO/PDRK. With the addition of only $S-1$ matrix elements, we can replace the matrix element $j_x$ by a sum over $S$ matrix elements $j_{sx}$, where $S$ is the number of species. 
To do this, we modified the BO/PDRK matrix equations in Eq. (\ref{eq:JmEnew}), i.e., 
\begin{eqnarray}
 \left\{ \begin{array}{ccc}
\omega j_{x}=b_{11}\delta E_x+b_{12}\delta E_y+b_{13}\delta E_z,\\
\omega j_{y}=b_{21}\delta E_x+b_{22}\delta E_y+b_{23}\delta E_z,\\
\omega j_{z}=b_{31}\delta E_x+b_{32}\delta E_y+b_{33}\delta E_z,
\end{array}\right.
\end{eqnarray}
as
\begin{eqnarray}
 \left\{ \begin{array}{ccc}
\omega j_{sx}=b_{s11}\delta E_x+b_{s12}\delta E_y+b_{s13}\delta E_z,\\
\omega j_{sy}=b_{s21}\delta E_x+b_{s22}\delta E_y+b_{s23}\delta E_z,\\
\omega j_{sz}=b_{s31}\delta E_x+b_{s32}\delta E_y+b_{s33}\delta E_z.
\end{array}\right.
\end{eqnarray}
This separation can directly give $j_{sx,y,z}$ from the BO/PDRK matrix, which then yields $\delta J_{sx,y,z}$ through the following equations
\begin{eqnarray}
 \left\{ \begin{array}{ccc}
\delta J_{sx}&=&(j_{sx}+\sum_{nj}^{s=m}v_{snjx}+\sum_{j\sigma}^{s=u}v_{sj\sigma x})/(i\epsilon_0),\\
\delta J_{sy}&=&(j_{sy}+\sum_{nj}^{s=m}v_{snjy}+\sum_{j\sigma}^{s=u}v_{sj\sigma y})/(i\epsilon_0),\\
\delta J_{sz}&=&(j_{sz}+\sum_{nj}^{s=m}v_{snjz}+\sum_{j\sigma}^{s=u}v_{sj\sigma z})/(i\epsilon_0).
\end{array}\right.
\end{eqnarray}

The updated matrix equations of BO, i.e., Eq.(132) of Ref.\cite{Xie2019}, become
\begin{equation}\label{eq:BOkemnew}
    \left\{ \begin{array}{ccc}
    \omega v_{snjx}^{s=m} &=& c_{snj} v_{snjx} + b_{snj11} \delta E_x + b_{snj12} \delta E_y + b_{snj13} \delta E_z, \\
    \omega v_{sj\sigma x}^{s=u} &=&  c_{sj\sigma } v_{sj\sigma x} + b_{sj\sigma 11} \delta E_x + b_{sj\sigma 12} \delta E_y + b_{sj\sigma 13} \delta E_z, \\
    \omega j_{sx} &=&  b_{s11} \delta E_x + b_{s12} \delta E_y + b_{s13} \delta E_z, \\
    i \delta J_x\epsilon_0 &=& j_x+\sum_{snj}^{s=m}v_{snjx}+ \sum_{sj\sigma }^{s=u}v_{sj\sigma x}, \\
    \omega v_{snjy}^{s=m} &=& c_{snj} v_{snjy} + b_{snj21} \delta E_x + b_{snj22} \delta E_y + b_{snj23} \delta E_z, \\
    \omega v_{sj\sigma y}^{s=u} &=& c_{sj\sigma } v_{sj\sigma y} + b_{sj\sigma 21} \delta E_x + b_{sj\sigma 22} \delta E_y + b_{sj\sigma 23} \delta E_z, \\
    \omega j_{sy} &=&  b_{s21} \delta E_x + b_{s22} \delta E_y + b_{s23} \delta E_z, \\
    i\delta J_y/\epsilon_0 &=& j_y+\sum_{snj}^{s=m}v_{snjy}+ \sum_{sj\sigma }^{s=u}v_{sj\sigma y}, \\
    \omega v_{snjz}^{s=m} &=& c_{snj} v_{snjz} + b_{snj31} \delta E_x + b_{snj32} \delta E_y + b_{snj33} \delta E_z, \\
    \omega v_{sj\sigma z}^{s=u} &=& c_{sj\sigma } v_{sj\sigma z} + b_{sj\sigma 31} \delta E_x + b_{sj\sigma 32} \delta E_y + b_{sj\sigma 33} \delta E_z, \\
    \omega j_{sz} &=&  b_{s31} \delta E_x + b_{s32} \delta E_y + b_{s33} \delta E_z, \\
    i\delta J_z/\epsilon_0 &=& j_z+\sum_{snj}^{s=m}v_{snjz}+\sum_{sj\sigma }^{s=u}v_{sj\sigma z}, \\
    \omega \delta E_x &=& c^2k_z \delta B_y-i\delta J_x/\epsilon_0, \\
    \omega \delta E_y &=& -c^2k_z \delta B_x +c^2k_x \delta B_z-i\delta J_y/\epsilon_0,\\
    \omega \delta E_z &=& -c^2k_x \delta B_y-i\delta J_z/\epsilon_0, \\
    \omega \delta B_x &=& -k_z \delta E_y, \\
    \omega \delta B_y &=& k_z \delta E_x - k_x \delta E_z, \\
    \omega \delta B_z &=& k_x \delta  E_y,
    \end{array}\right.
\end{equation}
which yields a sparse matrix eigenvalue problem $\omega{\bm X}={\bm M}({\bm k})\cdot{\bm X}$. The symbols such as $v_{snjx}$, $j_{sx,y,z}$ and $\delta J_{x,y,z}$ used here are analogous to the perturbed velocity and current density in fluid derivations of plasma waves. The elements of the eigenvector
$(\delta E_x, \delta E_y, \delta E_z, \delta B_x, \delta B_y,\delta  B_z)$ represent the perturbed electric and magnetic fields. Thus, all variables of one plasma wave mode can be obtained in a straightforward manner. In addition, the dimension of the matrix is $N_N=3\times (N_{SmNJ}+N_{SuJ}+S)+6=3\times
\{[S_m\times(2\times N+1)+S_u\times2]\times J+S\}+6$, where $S_m$ and $S_u$ are the numbers of magnetized and unmagnetized species, respectively, $S=S_m+S_u$, $N$ is the number of harmonics retained for magnetized species, and $J$ is the order of the $J$-pole expansion used for calculation of the plasma dispersion $Z$ function.

\subsection{Benchmark strategies}


In order to test whether the values of $(\delta {\bm E},\delta {\bm B},\delta{\bm J},\delta{\bm J}_s,\delta n_s,\delta{\bm v}_s)$ calculated by BO are correct, we do the following benchmarks:
\begin{itemize}
\item (1)	Use $\omega\delta {\bm B}={\bm k}\times\delta {\bm E}$ to check $\delta {\bm B}$  and $\delta {\bm J}=\frac{1}{\mu_0}i{\bm k}\times\delta {\bm B}+i\epsilon_0\omega\delta {\bm E}$  to check $\delta {\bm J}$.
\item (2)	Use $\delta {\bm J}_s={\bm\sigma}_s\cdot\delta {\bm E}$ to calculate $\delta {\bm J}_s$ and $\delta {\bm J}=\sum_s\delta {\bm J}_s$ to calculate $\delta {\bm J}$, and compare this with $\delta {\bm J}$ from (1).
\item (3)	Compare $\delta {\bm J}_s$ and $\delta {\bm J}$ in (2) with the analogous quantities calculated in the new version of BO/PDRK using $j_{sx,y,z}$ and $v_{snj}$.
\item (4)	Write out 3-by-3 tensors ${\bm \sigma}_s$, ${\bm Q}_s$, ${\bm\sigma}$, ${\bm Q}$, ${\bm K}$ and ${\bm D}(\omega,{\bm k})$ and verify that $|{\bm D}(\omega,{\bm k})|=0$.
\item (5)	Compare values of $\delta {\bm J}_s$, $\delta {\bm J}$, $\delta n_s$ and $\delta{\bm v}_s$ with those calculated using jWHAMP and the multi-fluid solver PDRF.
\end{itemize}
Using the new version of BO, we have checked the above (1)-(4) in several test cases and have identified a good consistence between these two methods proposed in Subsection 2.2. In following Section, we will present benchmark results in the above (5). 

\section{Benchmark and comparing with multi-fluid plasma model}\label{sec:bench}

\subsection{Benchmark with jWHAMP}

\begin{figure}
\centering
\includegraphics[width=16cm]{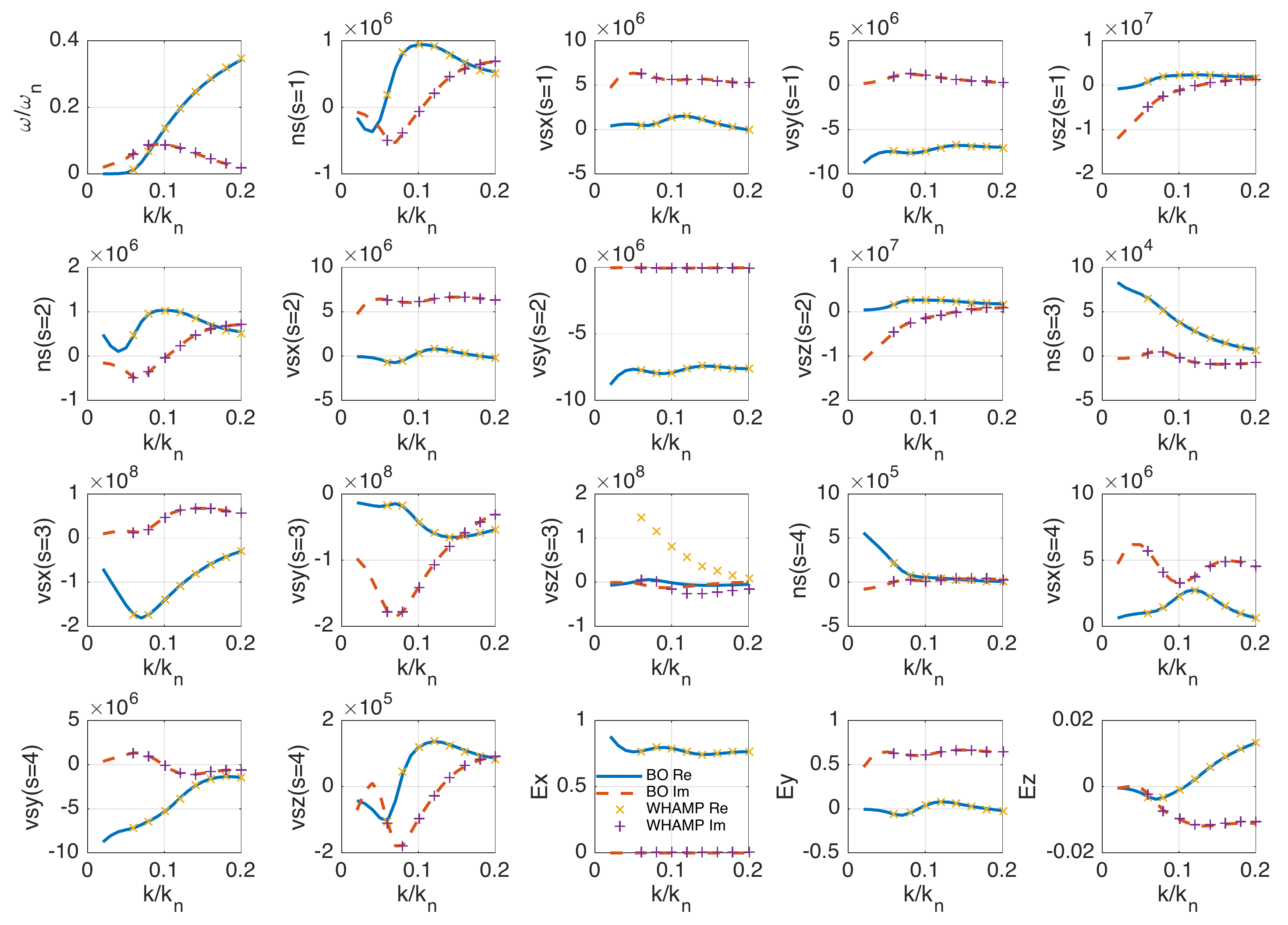}\\
\caption{Comparison of results from BO and jWHAMP with parameters case\#1. All of the quantities calculated by these two codes agree well except for $\delta v_{z}(s=3)$. BO contains the effect relating to $v_{dsz}$ for species 3, whereas $v_{dsz}\neq0$ was not taken into account in the current version of jWHAMP.}\label{fig:bo_jwhamp1}
\end{figure}

jWHAMP is Dartmouth College’s java extension of the WHAMP electromagnetic dispersion code \cite{Ronnmark1982}, which export a number of polarization quantities such as $(\delta{\bm E},\delta{\bm B},\delta n_s,\delta{\bm v}_s)$. However, in jWHAMP, the drift velocity, ${\bm v}_{ds}$, was not taken into account when calculating $\delta{\bm v}_s$. To test the greatest number of features of a kinetic calculation, we consider a case (case\#1) where the plasma consists of four species. We also consider both parallel and perpendicular components of the wave vector in case\#1.
The input species parameters for this case (specified in the `bo.in' input file) are
\begin{verbatim}
qs(e)   ms(m_unit) ns(m^-3)  Tzs(eV)     Tps(eV)    vdsz/c
1       1          1e6       24.838e3    99.352e3   0.0  
-1      5.447e-4   1.11e6    24.838e3    24.838e3   0.0  
1       1          0.01e6    24.838e4    24.838e4   0.0727 
1       4          0.1e6     0.1e3       0.1e3      0.0       
\end{verbatim}
Here, we use the default normalization in BO: the mass $m_s$ is normalized to the proton mass $m_p$, $k_x$ and $k_z$ are normalized to $k_n=\omega_{ps1}/c$ and the frequency is normalized to $\omega_n=|\omega_{cs1}|$, where ``1'' indicates the first species (i.e., the proton component with $n_s=10^6$ m$^{-3}$).
The case\#1 contains both anisotropic temperature and parallel drift velocity effects, which would destabilize the Alfv\'en/ion-cyclotron mode wave as shown in Fig. \ref{fig:bo_jwhamp1} that presents the distributions of the real and imaginary parts of frequency $\omega$, $\delta n_s$, $\delta {\bm v}_{s}$ and $\delta {\bm E}$ as a function of $k_z$ under $k_x=0.05$ and $B_0=100$ nT.

For all quantities presented in Fig. \ref{fig:bo_jwhamp1}, both BO and jWHAMP give the same distributions except for the values of $\delta v_{z}(s=3)$. The reason is that the effect of the drift was not included in jWHAMP to calculate $\delta v_{z}$. If we ignore $v_{dsz}$ in Eq. (\ref{eq:calnv}), we find that $\delta v_{z}(s=3)$ from BO agrees with the value from jWHAMP. 


This benchmark indicates that the new version of BO can correctly give the perturbed density and velocity in case\#1. Moreover, the comparison between BO and jWHAMP shows that it needs to take into account ${\bm v}_{ds}$ for calculating the perturbed velocity.

\subsection{Comparing with multi-fluid plasma model}
  
Here we compare the results of BO with those of a multi-fluid model. In the cold plasma limit ($v_{zts},v_{\perp ts}\simeq0$ or $\omega\gg k_zv_{zts}$), the kinetic results and fluid results should be identical. Note that we have updated the multi-fluid plasma dispersion relation solver PDRF, and that it can be run in the new version of BO (see \ref{sec:pdrf}).

\begin{figure}
\centering
\includegraphics[width=16cm]{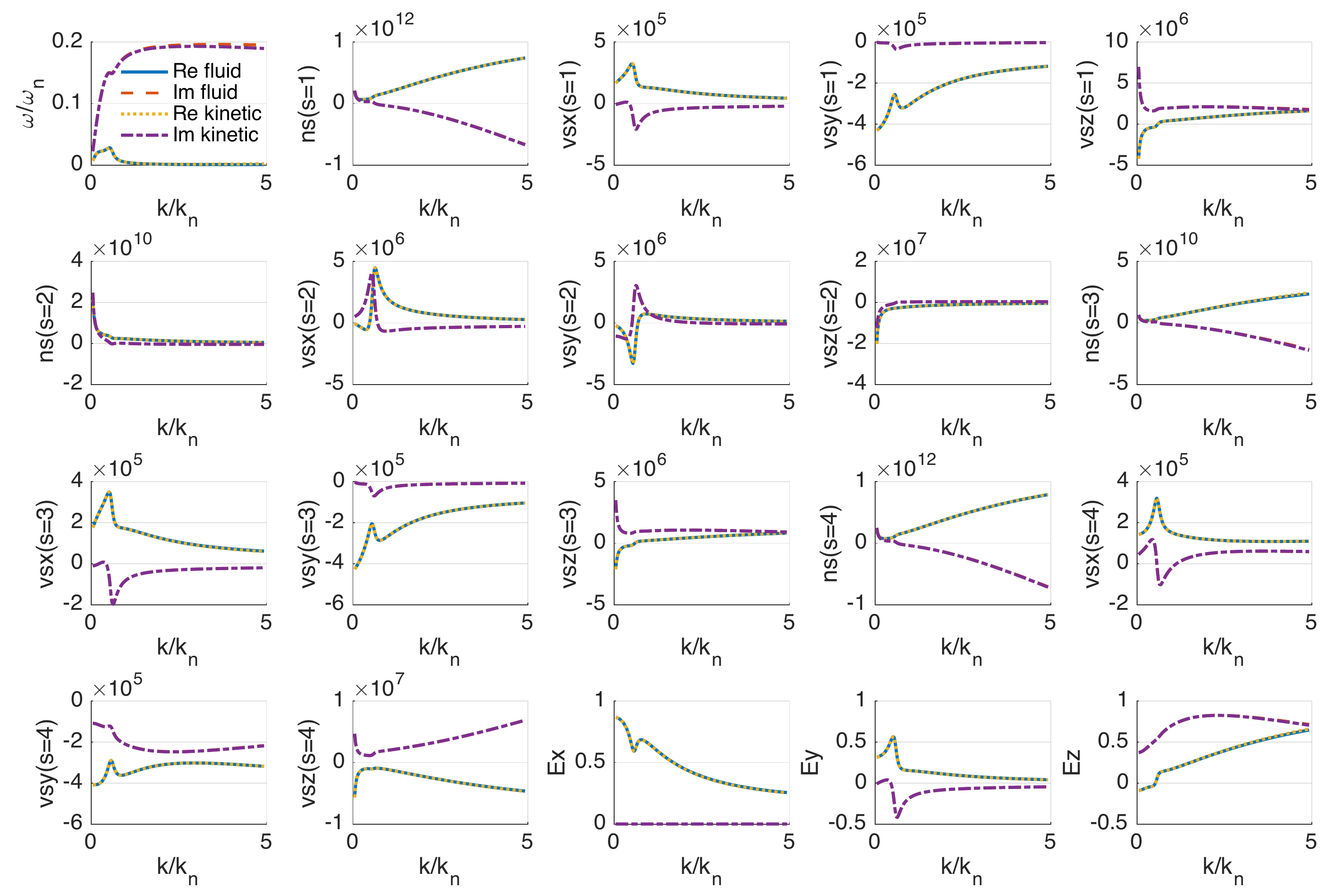}\\
\caption{Comparison of results from the BO kinetic and multi-fluid models for a cold plasma with parameters case\#2. The wave is the most unstable mode with the wave normal angle $\theta=10^\circ$. All quantities calculated using these two models agree well.}\label{fig:coldbeam_iem13_jpl=2}
\end{figure}

Fig. \ref{fig:coldbeam_iem13_jpl=2} compares results from the BO kinetic and fluid models for a cold four species plasma where $B_0=2000$ nT. The results are shown as a function of $k$ for the most unstable mode wave which has the wave normal angle $\theta=10^\circ$.
The input parameters for this case (case\#2) are
\begin{verbatim}
qs(e)   ms(m_unit) ns(m^-3)  Tzs(eV)     Tps(eV)   vdsz/c    vdsx/c    vdsy/c
1       1          0.8e10    1.0e-1      1.0e-1      0.0       0.0       0.0
1       1          0.1e10    1.0e-1      1.0e-1      2.0e-3    3.0e-3    1.0e-3
2       4          0.05e10   1.0e-1      1.0e-1      0.0       0.0       0.0
-1      5.447e-4   1.0e10    1.0e-1      1.0e-1      2.0e-4    3.0e-4    1.0e-4
\end{verbatim}
We consider both parallel and perpendicular beams in case\#2, and the default normalization is used. Fig. \ref{fig:coldbeam_iem13_jpl=2} shows that both kinetic and fluid models give the nearly same results. For large $k$ ($\sim5k_n$), there is a slight difference between $\omega_i$ for the the kinetic and fluid models, due to Landau damping in the kinetic model. If we decrease the temperature to $0.01$ eV, the deviation nearly vanishes. These results indicate that the equations in Sec.\ref{sec:eqn} and there implementation in BO are correct, even including perpendicular beams.



\begin{figure}
\centering
\includegraphics[width=15cm]{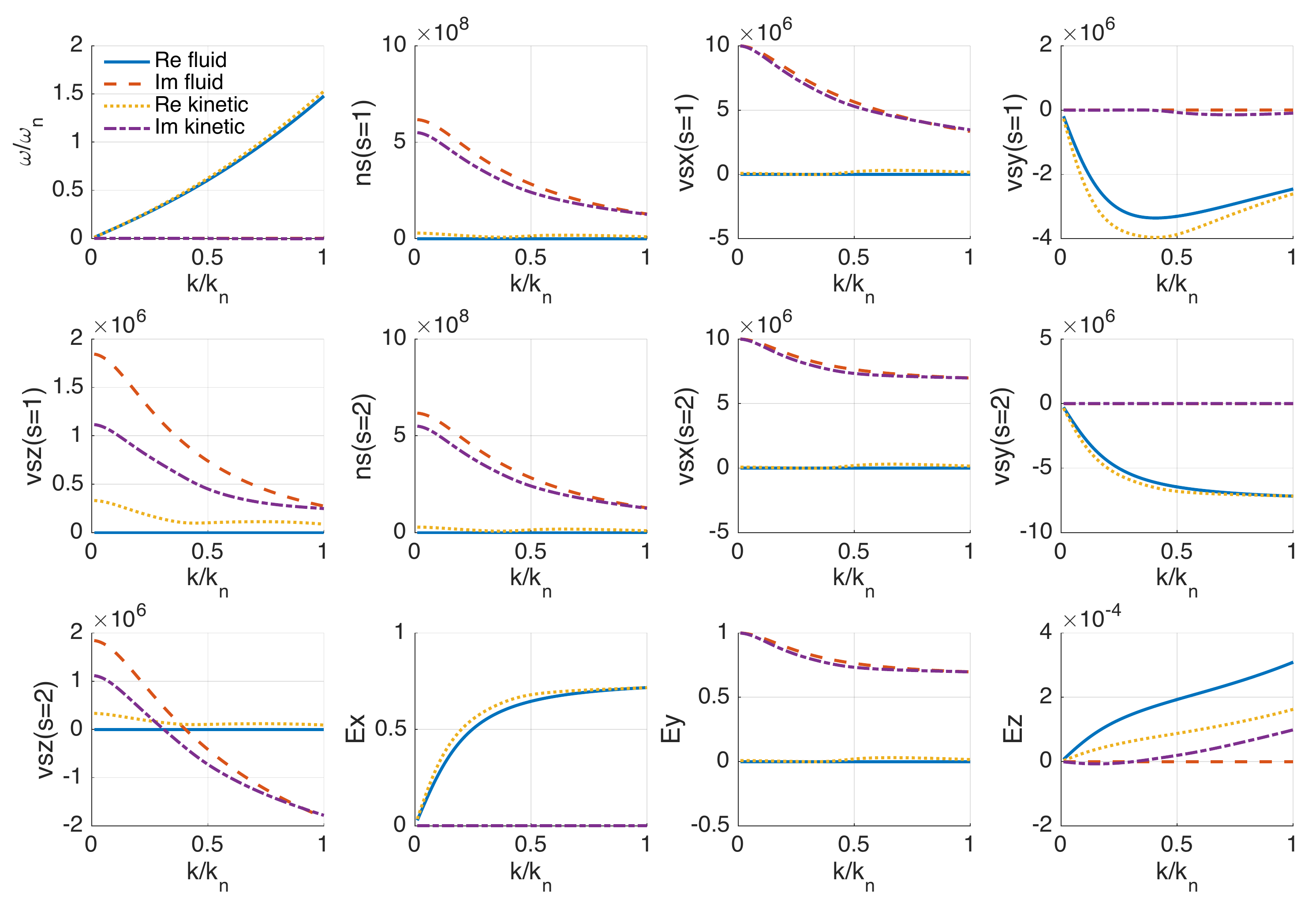}\\
\caption{Comparison of BO kinetic and fluid results for the fast-magnetosonic/whistler mode wave with parameters case\#3.}\label{fig:vA_theta30_branch1}
\end{figure}

\begin{figure}
\centering
\includegraphics[width=15cm]{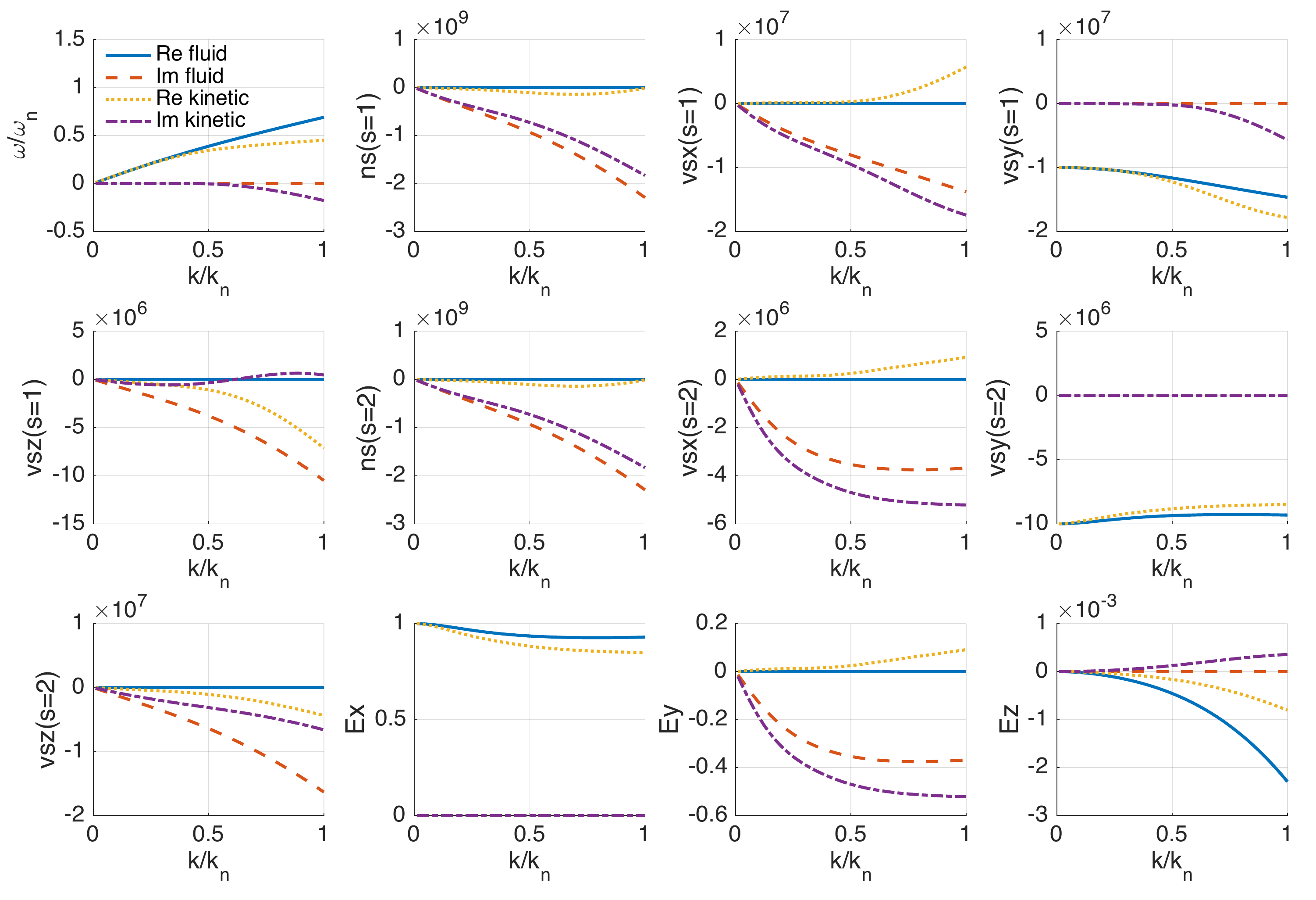}\\
\caption{Comparison of BO kinetic and fluid results for the Alfv\'en/ion-cyclotron mode wave with parameters case\#3. This wave is strongly damped at large $k$ in the kinetic model due to wave-particle interactions.}\label{fig:vA_theta30_branch2}
\end{figure}

We further compare results from the fluid and kinetic models for a warm plasma with isotropic pressure. The species parameters for this case  (case\#3) are
\begin{verbatim}
qs(e)   ms(m_unit) ns(m^-3)  Tzs(eV)     Tps(eV)
1       1          0.36e8    24.838e1    24.838e1
-1      5.447e-4   0.36e8    1.0         1.0
\end{verbatim} 
In order to have the consistent sound speed in both kinetic and fluid models, i.e., $c^2_s\simeq v^2_{ts}$, we choose the default adiabatic pressure closure with adiabatic coefficients $\gamma_{\parallel s} = \gamma_{\perp s}=2.0$. We also use $B_0=100$ nT and $\theta=30^\circ$. Figs. \ref{fig:vA_theta30_branch1} and \ref{fig:vA_theta30_branch2} give the results of the fast-magnetosonic/whistler mode and the Alfv\'en/ion-cyclotron mode, respectively. 
Since the kinetic wave-particle interactions considerably enhance in the warm plasma, the kinetic results would be different from the fluid results. Fig.\ref{fig:vA_theta30_branch1} shows that although the wave frequency from the kinetic and fluid models is almost the same, the quantities $(\delta{\bm E},\delta{\bm B},\delta n_s,\delta{\bm v}_s)$ have much larger deviations. Fig.\ref{fig:vA_theta30_branch2} shows that both the wave frequency and polarization have significant differences. 


\section{Summary and discussions}\label{sec:summ}

In this paper, we describe the updated BO plasma wave dispersion relation solver that can be used for both kinetic and fluid plasma models. 
We extend the kinetic version to obtain density and velocity perturbations for each species. In the cold plasma limit, the kinetic model yields results of $({\bm k},\omega,\delta{\bm E},\delta{\bm B},\delta{\bm J},\delta{n}_s,\delta{v}_s)$ quite similar to those from the multi-fluid model, even for zeroth order drift beams in arbitrary directions. In a warm plasma, Landau and cyclotron wave-particle resonance effects can alter the wave frequency and the polarization, which induce the difference between the kinetic and fluid models. The extensive set of polarization quantities calculated by the updated BO (see Appendix~B) could be useful for identifying and characterizing plasma waves and instabilities in space plasmas. 


{\it Acknowledgments}
Work at Dartmouth College was supported by NASA grant 80NSSC19K0270. 

\appendix

\section{Reduced version of multi-fluid plasma dispersion relation solver PDRF}\label{sec:pdrf}

To make the PDRF code more amenable to comparison with BO-K/PDRK, we simplify the original version of PDRF by removing density inhomogeneity, relativistic effects and collisions, and make it available as BO-F in BO code. Drifts in arbitrary directions and pressure anisotropy are retained. Some typos in Ref.\cite{Xie2014} are also corrected here.

We consider a multi-fluid plasma in an external magnetic field $\bm
B_0=(0,0,B_0)$. The zero-th order flow velocity of the fluid component $s$ is $\bm
v_{ds}=(v_{dsx},v_{dsy},v_{dsz})$. The species densities and temperatures are homogeneous, i.e., gradient effects are ignored, and the wave
vector is assumed to be
$\bm{k}=(k_x,0,k_z)=(k\sin\theta,0,k\cos\theta)$. 

We start with the mulit-fluid equations
\begin{subequations} \label{eq:fpeq}
\begin{eqnarray}
  & \partial_t n_s = -\nabla\cdot(n_s\bm v_s),\\
  & \partial_t \bm v_s = -\bm v_s\cdot \nabla\bm v_s+\frac{q_s}{m_s}(\bm E+\bm v_s\times \bm B)-\frac{\nabla\cdot\bm P_s}{\rho_s},\\
  & \partial_t \bm E = c^2\nabla\times\bm B - \bm J/\epsilon_0,\\
  & \partial_t \bm B = -\nabla\times\bm E,
\end{eqnarray}
\end{subequations}
where we ignore the relativistic effects, and
\begin{subequations} \label{eq:fpeq2}
\begin{eqnarray}
  & \bm J = \sum_sq_sn_s\bm v_s, \\  
  & d_t(P_{\parallel,\perp s}\cdot\rho_s^{-\gamma_{\parallel,\perp s}}) = 0,
\end{eqnarray}
\end{subequations}
where the mass density is $\rho_{s}\equiv m_sn_{s}$, and the speed of light is $c=1/\sqrt{\mu_0\epsilon_0}$.  In the above equations, 
we have used adiabatic model for pressure closure, with $\gamma_{\parallel,\perp s}$ being the parallel and perpendicular exponents. 
Furthermore,
$P_{\parallel,\perp}=nk_BT_{\parallel,\perp}$, $\bm
P=P_{\parallel}\hat{\bm b}\hat{\bm b}+P_{\perp}(\bm I-\hat{\bm
b}\hat{\bm b})$ and $\hat{\bm b}=\bm B/B$. 
Different anisotropic pressure closures will yield different results.
Usually, one take $\gamma_{\parallel}=\gamma_{\perp}=5/3$. However, we find $\gamma_{\parallel}=\gamma_{\perp}=2$ would yield closer results to those of the kinetic model. If not specified by the user, $\gamma_{\parallel}=\gamma_{\perp}=2$ are the default settings. 


After linearizing, (\ref{eq:fpeq2}) becomes
\begin{subequations} \label{eq:fpeq2lin}
\begin{eqnarray}
  & \delta\bm J = \sum_sq_s(n_{s0} \delta\bm v_{s}+ \delta n_{s}\bm v_{ds}), \\
  &  \delta P_{\parallel,\perp s} =P_{\parallel,\perp s0}\gamma_{\parallel,\perp s}\delta n_s/n_{s0}= c^2_{\parallel,\perp s} m_s\delta n_{s},
\end{eqnarray}
\end{subequations}
where $c^2_{\parallel,\perp s}\equiv\gamma_{\parallel,\perp s} P_{\parallel,\perp s0}/\rho_{s0}$ and $\bm P_{s0}=n_{s0}k_B\bm
T_{s0}$. We also define $c_{As}\equiv B_0^2/(\mu_0\rho_{s0})$.
We have
\begin{equation} \label{eq:gradP}
   \nabla\cdot\delta \bm P_{s} = (ik_x,0,ik_z)\cdot\left[\begin{array}{ccc}
    \delta P_{\perp s} & 0 & \Delta_s \delta B_{x}\\
    0 & \delta P_{\perp s} & \Delta_s \delta B_{y}\\
    \Delta_s \delta B_{x} & \Delta_s \delta B_{y} & \delta P_{\parallel s}
    \end{array}
 \right]=\left(
 \begin{array}{c}ik_x\delta P_{\perp s}+ik_z\Delta_s \delta B_{x}\\
 ik_z\Delta_s \delta B_{y}\\
 ik_x\Delta_s \delta B_{x}+ik_z\delta P_{\parallel s}
 \end{array}\right)^T,
\end{equation}
where $\Delta_s\equiv(P_{\parallel s0}-P_{\perp s0})/B_0$ and
$\beta_{\parallel,\perp s}=2\mu_0P_{\parallel,\perp s}/B_0^2$. The
off-diagonal terms coming from the tensor rotation from $\hat{\bm b}_0$ to $\hat{\bm
b}$ are related to energy exchange and are important for the
anisotropic instabilities. 

The linearized version of (\ref{eq:fpeq}) with
$f=f_0+\delta fe^{i\bm{k\cdot r}-i\omega t}$, $\delta f\ll f_0$ is equivalent to a matrix
eigenvalue problem
\begin{equation}\label{eq:eig}
    \omega \bm
X=\bm M\bm X,
\end{equation}
where $\omega$ is the eigenvalue and $\bm X$ is the corresponding
eigenvector containing polarization information for the eigenvectors. 
Accordingly, we have $\bm X=(\delta n_{s},\delta v_{sx},\delta v_{sy},\delta v_{sz},\delta E_{x},\delta E_{y},\delta E_{z},\delta B_{x},\delta B_{y},\delta B_{z})^T$, and 
the matrix 
\begin{equation}\label{eq:fpdrM}
\bm M=\left[\begin{array}{cc}
\left\{\begin{array}{cccc}
\bm k\cdot\bm v_{ds} & k_xn_{s0} & 0 & k_zn_{s0} \\
\frac{k_xc_{\perp s}^2}{n_{s0}} & \bm k\cdot\bm v_{ds} & i\omega_{cs} & 0\\
0 & -i\omega_{cs} & \bm k\cdot\bm v_{ds} & 0  \\
 \frac{k_zc_{\parallel s}^2}{n_{s0}} & 0 & 0 & \bm k\cdot\bm v_{ds}
    \end{array}
 \right\}
 & \left\{\begin{array}{cccccc}
 0 & 0 & 0 & 0 & 0 & 0\\
 i\frac{q_s}{m_s} & 0 & 0 & \frac{k_z\Delta_s}{m_sn_{s0}} & -i\frac{q_sv_{dsz}}{m_s} & i\frac{q_sv_{dsy}}{m_s}\\
 0 & i\frac{q_s}{m_s} & 0 & i\frac{q_sv_{dsz}}{m_s} & \frac{k_z\Delta_s}{m_sn_{s0}} & -i\frac{q_sv_{dsx}}{m_s}\\
  0 & 0 & i\frac{q_s}{m_s} & -i\frac{q_sv_{dsy}}{m_s}+\frac{k_x\Delta_s}{m_sn_{s0}} & i\frac{q_sv_{dsx}}{m_s} & 0
  \end{array}
 \right\}
 \\
 
\left\{\begin{array}{cccc}
-i\frac{q_sv_{dsx}}{\epsilon_0} & -i\frac{q_sn_{s0}}{\epsilon_0} & 0 & 0 \\
-i\frac{q_sv_{dsy}}{\epsilon_0} & 0 & -i\frac{q_sn_{s0}}{\epsilon_0} & 0 \\
-i\frac{q_sv_{dsz}}{\epsilon_0} & 0 & 0 & -i\frac{q_sn_{s0}}{\epsilon_0}\\
0 & 0 & 0 & 0  \\
0 & 0 & 0 & 0  \\
0 & 0 & 0 & 0  
    \end{array}
 \right\}
 & \begin{array}{cccccc}
  0 & 0 & 0 & 0 & k_zc^2 & 0\\
  0 & 0 & 0 & -k_zc^2 & 0 & k_xc^2\\
  0 & 0 & 0 & 0 & -k_xc^2 & 0\\
  0 & -k_z & 0 & 0 & 0 & 0\\
  k_z & 0 & -k_x & 0 & 0 & 0\\
  0 & k_x & 0 & 0 & 0 & 0
  \end{array}

    \end{array}
 \right],
\end{equation}
where that the elements between `$\{$' and `$\}$' means each species $s$ has its own matrix elements, $\omega_{cs}=q_sB_0/m_s$, $q_e=-e$,
$\omega_{ps}^2=n_{s0}q_s^2/(\epsilon_0m_s)$, and $\bm k\cdot\bm v_{ds}=k_xv_{dsx}+k_zv_{dsz}$. 
For a plasma containing $S$ species, the dimension of $\bm M$ is $(4S+6)\times(4S+6)$. 
If we define the thermal velocity $v_{\parallel,\perp  ts}=\sqrt{2k_BT_{\parallel,\perp s}/m_s}$ as in the kinetic version of BO\cite{Xie2019}, we can have $c_{\parallel,\perp s}=\sqrt{\gamma_{\parallel,\perp s}/2}v_{\parallel,\perp  ts}$, or the temperature $k_BT_{\parallel,\perp s0}=m_{s}c^2_{\parallel,\perp s}/\gamma_{\parallel,\perp s} $. 

We can also use some other pressure closures. For example, the double-polytropic laws for pressure closure
\begin{subequations} 
\begin{eqnarray}\label{eq:pclosure2}
  & d_t(P_{\parallel s}\cdot\rho_s^{-\gamma_{\parallel s}}\cdot B^{\gamma_{\parallel s}-1}) = 0, \\
  & d_t(P_{\perp s}\cdot\rho_s^{-1}\cdot B^{-\gamma_{\perp s}+1}) = 0,
\end{eqnarray}
\end{subequations} 
with $\gamma_{\parallel}$ and $\gamma_{\perp s}$ being the parallel and perpendicular polytrope exponents had been used previously in space plasma studies, c.f., Ref.\cite{Hau1993}.  Note that $\gamma_{\parallel}=3$ and  $\gamma_{\perp}=2$ yield the CGL relations\cite{Chew1956}, whereas $\gamma_{\parallel}=\gamma_{\perp}=1$ yields isothermal behavior.  For this pressure closure, we have
\begin{subequations} 
\begin{eqnarray}
  &  \delta P_{\parallel s} =P_{\parallel s0}[\gamma_{\parallel s}\delta n_s/n_{s0}-(\gamma_{\parallel s}-1)\delta B_z/B_0]= c^2_{\parallel s} m_s\delta n_{s}-P_{\parallel s0}(\gamma_{\parallel s}-1)\delta B_z/B_0,\\
  &  \delta P_{\perp s} =P_{\perp s0}[\delta n_s/n_{s0}+(\gamma_{\perp s}-1)\delta B_z/B_0]= c^2_{\perp s} m_s\delta n_{s}+P_{\perp s0}(\gamma_{\perp s}-1)\delta B_z/B_0,
\end{eqnarray}
\end{subequations} 
where $c^2_{\parallel s}\equiv\gamma_{\parallel s} P_{\parallel s0}/\rho_{s0}$ and $c^2_{\perp s}\equiv P_{\perp s0}/\rho_{s0}$ (here is different from the $c^2_{\perp s}\equiv \gamma_{\perp}P_{\perp s0}/\rho_{s0}$ in Ref.\cite{Hau1993}), and hence the matrix elements $M_{\delta v_{sx},\delta n_s}$, $M_{\delta v_{sx},\delta B_z}$ and $M_{\delta v_{sz},\delta B_z}$ in ${\bm M}$ would be modified accordingly. Similarly to the adiabatic pressure closure case, we can have $c_{\parallel s}=\sqrt{\gamma_{\parallel s}/2}v_{\parallel  ts}$ and $c_{\perp s}=\sqrt{1/2}v_{\perp  ts}$, or the temperature $k_BT_{\parallel s0}=m_{s}c^2_{\parallel s}/\gamma_{\parallel s} $ and $k_BT_{\perp s0}=m_{s}c^2_{\perp s} $. 

In BO, because of the limitations of the pressure closure, we only use the above fluid version to get a rough description of the waves and instabilities and for comparison with the kinetic version. It is especially useful for studying cold plasma waves and beam modes, in which case the pressure closure is not important. The fluid closure has many limitations and leads to some un-physical results. For example, in the double-polytropic CGL case ($\gamma_{\parallel}=3$, $\gamma_{\perp} = 2$), the waves can be unstable even when $P_{\parallel} = P_{\perp s0}=P_{\parallel s0}$ because $c^2_{\parallel s}\neq c^2_{\perp s}$. The high beta anisotropic firehose and mirror mode instabilities are also difficult to calculate accurately from fluid model. For accurate results with finite pressure, we recommend the kinetic version of BO.


\section{Polarizations in BO}
For given real $k_x$, $k_z$ and corresponding complex $\omega$,  we find the complex quantities $\delta E_x$, $\delta E_y$, $\delta E_z$, $\delta B_x$, $\delta B_y$, $\delta B_z$, $\delta J_x$, $\delta J_y$, $\delta J_z$, $\delta J_{sx}$, $\delta J_{sy}$, $\delta J_{sz}$, $\delta n_{s}$, $\delta v_{sx}$, $\delta v_{sy}$, $\delta v_{sz}$. We list the comprehensive  polarization quantities calculated in the new version of BO code, and summarize them in Table \ref{tab:pola}. Note that for a given $(k_x,k_z)$, there exist multiple branches corresponding to different eigenmodes $\omega$, and each branch has its unique polarization. 

\begin{table}[htbp]
\caption{List of the polarization quantities in BO. The numbers before each polarizations are the default indexes of them used in the BO code. We set npf=50 and nps=50 by default. For example, since npf+nps*(s-1)+4 is the $s$-th perturbed density, which means 50+50*(2-1)+4=104 is the default index of the perturbed density for the 2nd species.}\label{tab:pola}
\begin{center}
{\footnotesize
\begin{tabular}{l c | l c}
\hline\hline
1. electric field in $x$-direction ($V/m$) &  $\delta E_x$ & 2. electric field in $y$-direction  ($V/m$) & $\delta E_y$ \\

3. electric field in $z$-direction  ($V/m$) & $\delta E_z$ &4. magnetic field in $x$-direction ($T$) &  $\delta B_x$ \\

5. magnetic field in $y$-direction ($T$) &  $\delta B_y$ & 6. magnetic field in $z$-direction ($T$) &  $\delta B_z$\\

7. electric field energy density ($J/m^3$) & $\delta U_E$ & 8. Magnetic field energy density ($J/m^3$) &  $\delta U_B$ \\

9. fraction of field energy in the electric field & $\frac{\delta U_E}{(\delta U_E+\delta U_B)}$ & 10. fraction of electric field energy in $\delta E_x$ &$\frac{|\delta E_x|^2}{|\delta E|^2}$\\

11. fraction of electric field energy in $\delta E_y$ & $\frac{|\delta E_y|^2}{|\delta E|^2}$ & 12. fraction of electric field energy in $\delta E_z$ &$\frac{|\delta E_z|^2}{|\delta E|^2}$\\
13. a measure of how electrostatic is& $\frac{| {\bm k} \cdot \delta {\bm E} |^2 }{ ( k^2  |\delta E|^2 )}$ &  14. another measure
        of how electrostatic is & $\frac{| {\bm k} \cdot \delta {\bm E} |^2 }{  | {\bm k} \cdot \delta {\bm E} |^2 + | {\bm k} \times \delta {\bm E} |^2 }$ \\
15. fraction of magnetic field energy in $\delta B_x$ &$\frac{|\delta B_x|^2}{|\delta B|^2}$ &  16. fraction of magnetic field energy in $\delta B_y$ &$\frac{|\delta B_y|^2}{|\delta B|^2}$\\
17. fraction of magnetic field energy in $\delta B_z$ &$\frac{|\delta B_z|^2}{|\delta B|^2}$ &  18. magnetic polarization ellipticity & $\epsilon_B$\\
19. angle of major axis of magnetic ellipse  & $\theta_B$ & 20. magnetic polarization ratio, $|\alpha_B|$ & $\frac{|\delta B_y|}{|\delta B_x|}$ \\
21. angle of magnetic polarization ratio, $\phi_B$ & $\arg\Big(\frac{\delta B_y}{\delta B_x}\Big)$  &  22. wave group velocity in $x$-direction ($m/s$) & $v_{gx}$\\
23. wave group velocity in $z$-direction ($m/s$)& $v_{gz}$ & 24. spatial growth rate in $x$-direction, $S_x$ & $\frac{\omega_i}{v_{gx}}$\\
25. spatial growth rate in $z$-direction ($m^{-1}$), $S_z$ & $\frac{\omega_i}{v_{gz}}$ &  26. total spatial growth rate ($m^{-1}$), $S$ & $\frac{\omega_i}{|v_{g}|}$\\

27. refractive index & $|\bm n|$ &  28 to 36. dispersion tensor elements $i,j=1,2,3$ & $D_{ij}$\\

$\cdots$ & $\cdots$ &  $\cdots$ & $\cdots$\\

npf-6. current density in $x$-direction ($A/m^2$) & $\delta J_x$ &  npf-5. current density in $y$-direction ($A/m^2$) & $\delta J_y$\\
npf-4. current density in $z$-direction ($A/m^2$) & $\delta J_z$   & npf-3. perpendicular wave vector ($m^{-1}$) & $k_x$ \\
 npf-2.  parallel wave vector  ($m^{-1}$) &  $k_z$ & npf-1. wave vector  ($m^{-1}$) & $k$ \\
npf. wave frequency ($s^{-1}$)& $\omega$ &  npf+nps*(s-1)+1. $s$-th perturbed $x$-current ($A/m^2$) & $\delta J_{sx}$\\
npf+nps*(s-1)+2. $s$-th perturbed $y$-current ($A/m^2$) & $\delta J_{sy}$ &  npf+nps*(s-1)+3. $s$-th perturbed $z$-current ($A/m^2$) & $\delta J_{sz}$\\
npf+nps*(s-1)+4. $s$-th perturbed density ($m^{-3}$) & $\delta n_{s}$ &  npf+nps*(s-1)+5. $s$-th perturbed $x$-velocity ($m/s$) & $\delta v_{sx}$\\
npf+nps*(s-1)+6.  $s$-th perturbed $z$-velocity ($m/s$)  & $\delta v_{sy}$ &  npf+nps*(s-1)+7.  $s$-th perturbed $z$-velocity ($m/s$)  & $\delta v_{sz}$\\
npf+nps*(s-1)+8.  $s$-th species compressibility & $\frac{|\delta n_{s}/n_{s0}|^2}{|\delta B/B_0|^2}$ &  npf+nps*(s-1)+9.  $s$-th species Alfven-ratio & $\frac{|\delta v_{s}/v_{A}|^2}{|\delta B/B_0|^2}$ \\
npf+nps*(s-1)+10.  $s$-th parallel phase
 ratio & $\frac{Re[\delta n_s\cdot\delta B_z^*]}{|\delta n_s||\delta B_z|}$ &  npf+nps*(s-1)+11. $s$-th  kinetic energy fraction $x$  & $\frac{|\delta v_{sx}|^2}{|\delta v_s|^2}$\\
npf+nps*(s-1)+12. $s$-th  kinetic energy fraction $y$ & $\frac{|\delta v_{sy}|^2}{|\delta v_s|^2}$ &  npf+nps*(s-1)+13. $s$-th kinetic energy fraction  $z$  & $\frac{|\delta v_{sz}|^2}{|\delta v_s|^2}$\\
npf+nps*(s-1)+14. $s$-th kinetic energy fraction  $k$  & $\frac{|{\bm k}\cdot\delta {\bm v}_{s}|^2}{|k|^2|\delta v_s|^2}$ &  npf+nps*(s-1)+15. $s$-th perturbed velocity magnitude & $|\delta {\bm v}_s|$\\
npf+nps*(s-1)+16. $s$-th kinetic energy  & $\delta U_{Ks}$ &  npf+nps*(s-1)+17. $s$-th kinetic energy fraction  & $\frac{\delta U_{Ks}}{\delta U_{tot}}$ \\
npf+nps*(s-1)+18:26. $s$-th conductivity tensor elements & $\sigma_{s,ij}$ &  $\cdots$ & $\cdots$\\
$\cdots$ & $\cdots$ &  $\cdots$ & $\cdots$\\
\hline\hline
\end{tabular}
}
\end{center}
\label{tab:polalist}
\end{table}%

The linear polarizations can have arbitrary large magnitude. After we obtain the eigenvectors $\delta{\bm X}_0=[\delta E_{x0},\delta E_{y0},\delta E_{z0},\delta B_{x0},\cdots]$ in the code, the normalization of modes is done like this
\begin{equation}
\delta{\bm X}_1=\frac{\delta{\bm X}_0}{\delta E_{x0}},~~~\delta{\bm X}=\frac{\delta{\bm X}_1}{|\delta E_{x1}|^2+|\delta E_{y1}|^2+|\delta E_{z1}|^2},
\end{equation}
which causes $\delta E_x$ to be real and positive, and $|\delta E|=\sqrt{|\delta E_x|^2+|\delta E_y|^2+|\delta E_y|^2}=1$ V/m. This procedure will work except for some extreme cases for which $|\delta E_x|<10^{-16}|\delta E|$ in double precision calculations. Equations to calculate some other relevant field quantities are
\begin{eqnarray}
&|\delta E_x|^2=\delta E_x\cdot\delta E_x^*=[Re(\delta E_x)]^2+[Im(\delta E_x)]^2,~~~{\rm similar~ for~~} |\delta E_y|^2, |\delta E_z|^2, |\delta B_x|^2, |\delta B_y|^2, |\delta B_z|^2\\
&|\delta {\bm E}|^2=|\delta E|^2=|\delta E_x|^2+|\delta E_y|^2+|\delta E_z|^2, ~~\delta U_E=\frac{1}{2}\cdot\frac{1}{2}\epsilon_0 |\delta E|^2,\\
&|\delta {\bm B}|^2=|\delta B|^2=|\delta B_x|^2+|\delta B_y|^2+|\delta B_z|^2, ~~\delta U_B=\frac{1}{2}\cdot\frac{1}{2} \frac{|\delta B|^2}{\mu_0},\\
& | {\bm k} \cdot \delta {\bm E} |^2=|k_x\cdot\delta E_x+k_z\cdot\delta E_z|^2,\\
& | {\bm k} \times \delta {\bm E} |^2=|-k_z\cdot\delta E_y|^2+|k_z\cdot\delta E_x-k_x\cdot\delta E_z|^2+|k_x\cdot\delta E_y|^2,
\end{eqnarray}
where the asterisk denotes complex conjugation. The extra $\frac{1}{2}$ in $\delta U_E$  and $\delta U_B$ is for a time average. Note that usually $|\delta E_{x,y,z}|^2\neq\delta E_{x,y,z}^2$, because $|\delta E_{x,y,z}|^2$ is always real, whereas $\delta E_{x,y,z}^2$ is usually complex.
The total energy density $\delta U_{tot}=\delta U_E+\delta U_B+\delta U_K$, where
\begin{equation}
\delta U_K=\sum_s\delta U_{Ks},~~\delta U_{Ks}=\frac{1}{2}\cdot\frac{1}{2}m_sn_{s0}|\delta v_s|^2+\frac{1}{2}\cdot\frac{1}{2}m_s(v_{dsx}^2+v_{dsy}^2+v_{dsz}^2)|\delta n_s|.
\end{equation}

To get the magnetic ellipticity, we use
\begin{eqnarray}
\alpha_B\equiv\frac{\delta B_y}{\delta B_x}=|\alpha_B|e^{i\phi_B},~~~{\rm with~~} |\alpha_B|=\Big|\frac{\delta B_y}{\delta B_x}\Big|,~~\phi_B\equiv\arg\Big(\frac{\delta B_y}{\delta B_x}\Big),\\
\delta B_L=\delta B_x+i\delta B_y,~~~\delta B_R=\delta B_x-i\delta B_y,~~~\epsilon_B=\frac{|\delta B_R|-|\delta B_L|}{|\delta B_R|+|\delta B_L|}.
\end{eqnarray}
If $\epsilon_B = 1$, the wave is right hand circularly polarized, if $\epsilon_B = 0$, the wave is linearly polarized, and if $\epsilon_B = -1$, the wave is left-hand circularly polarized.

The quantity $\theta_B$ is the angle of major axis of the magnetic ellipse, and it should satisfy
\begin{eqnarray}
\theta_B=\tan^{-1}\Big[\frac{|\alpha_B|\cos(\phi_B+\varphi)}{\cos\varphi}\Big],
\end{eqnarray}
where $\varphi$ is the angle for the following quantity be at maximum
\begin{eqnarray}
f(\varphi)=\cos^2\varphi+|\alpha_B|^2\cos^2(\phi_B+\varphi),
\end{eqnarray}
i.e., its derivative vanishes
\begin{eqnarray}
\frac{\partial}{\partial\varphi}[\cos^2\varphi+|\alpha_B|^2\cos^2(\phi_B+\varphi)]=-2\sin(2\varphi)-2|\alpha_B|^2\sin[2(\phi_B+\varphi)]=0,\\
\frac{\partial^2}{\partial\varphi^2}[\cos^2\varphi+|\alpha_B|^2\cos^2(\phi_B+\varphi)]=-4\cos(2\varphi)-4|\alpha_B|^2\cos[2(\phi_B+\varphi)]<0,
\end{eqnarray}
which yields
\begin{eqnarray}
\varphi=\frac{1}{2}\cot^{-1}\Big[-\frac{(1/|\alpha_B|^2)+\cos(2\phi_B)}{\sin(2\phi_B)}\Big], ~~{\rm and~required~} \cos(2\varphi)+|\alpha_B|^2\cos[2(\phi_B+\varphi)]>0.
\end{eqnarray}
If $\cos(2\varphi)+|\alpha_B|^2\cos[2(\phi_B+\varphi)]<0$, we set $\varphi\to\varphi+\frac{\pi}{2}$ since the difference between major axis and minor axis is $\frac{\pi}{2}$.
Using above equations, we can obtain $|\alpha_B|$, $\phi_B$ and $\theta_B$. If $\phi_B=\pi/2$ or $\in(0,\pi)$, the wave is left-hand polarization; if $\phi_B=-\pi/2$ or $\in(-\pi,0)$, the wave is right-hand polarization; if $\phi_B\simeq0$, or $|\alpha_B|\gg1$ or $\ll1$, the wave is linear polarization; if $|\alpha_B|=1$, the wave is circular spolarization; other case, the wave is elliptical polarization. 



The species compressibility, parallel phase ratio and Alfven-ratio are calculated using definitions in Refs.\cite{Gary1992,Denton1998}. The refractive index $\bm n=\frac{c{\bm k}}{\omega}$. The phase velocity ${\bm v}_p=\frac{\omega}{\bm k}$, the group velocity ${\bm v}_g=\frac{\partial \omega}{\partial \bm k}$.  For $F(\omega,{\bm k})=0$, using the implicit function derivative formula, we have ${\bm v}_g=\frac{\partial \omega}{\partial \bm k}=-\frac{\partial F/\partial {\bm k} }{\partial F/\partial {\omega}}$, where $F(\omega,{\bm k})=|{\bm M}({\bm k})-\omega|$ or $F(\omega,{\bm k})=|D(\omega,{\bm k})|$. However, calculating this is very complicated. Thus, we can use numerical differentiation to calculate the group velocity; i.e., after we obtain a $\omega$ for a given $(k_x,k_z)$, we solve the dispersion relation with $(k_x+\Delta k_x,k_z)$ and $(k_x,k_z+\Delta k_z)$, which gives $\omega+\Delta\omega_x$ and  $\omega+\Delta\omega_z$. The group velocity is then $v_{gx}=\frac{\Delta\omega_x}{\Delta k_x}$ and $v_{gz}=\frac{\Delta\omega_z}{\Delta k_z}$. 

\section{Typos or bugs fixed in BO 1.0}
In BO version 1.0\cite{Xie2019}, some typos are found. In page 352, $T_{zs}=\frac{1}{2}k_Bm_sv_{zts}^2$ and $T_{\perp s\sigma}=\frac{1}{2}k_Bm_sv_{\perp ts\sigma}^2$ should be $k_BT_{zs}=\frac{1}{2}m_sv_{zts}^2$ and $k_BT_{\perp s\sigma}=\frac{1}{2}m_sv_{\perp ts\sigma}^2$.

In page 360,
\begin{itemize}
\item $P_{s32}^{m}=\sum_{n=-\infty}^{\infty}\int_{-\infty}^{\infty}\int_0^{\infty}  \frac{2\pi v'_\perp dv'_\perp dv''_\parallel}{(\omega_{sn}-k_z v''_\parallel )} \Pi_{s\sigma32}=\sum_{n=-\infty}^{\infty} \Big\{  [ n\omega_{cs} \frac{ A_{nbs\sigma}}{v_{\perp ts\sigma}^2}\frac{Z_1}{k_z}+A_{n0\sigma} \frac{k_z v_{zts}Z_2 }{v_{zts}^2}]v_{dsy}+ [ n\omega_{cs} \frac{ A_{nbs\sigma}}{v_{\perp ts\sigma}^2}\frac{Z_0}{k_zv_{zts}}+A_{n0\sigma} \frac{Z_1 }{v_{zts}^2}]v_{dsy}v_{dsz} +i [(n\omega_{cs} -i\nu_s )\frac{ B_{nbs\sigma}}{v_{\perp ts\sigma}}\frac{Z_1}{k_z}+B_{n0\sigma} \frac{  v_{\perp ts\sigma}Z_2}{v_{zts} }]+i v_{dsz} [(n\omega_{cs} -i\nu_s )\frac{ B_{nbs\sigma}}{v_{\perp ts\sigma}}\frac{Z_0}{k_zv_{zts}}+B_{n0\sigma} \frac{ v_{\perp ts\sigma}Z_1}{v_{zts}^2}]   \Big\} $.
\end{itemize}
should be
\begin{itemize}
\item $P_{s32}^{m}=\sum_{n=-\infty}^{\infty}\int_{-\infty}^{\infty}\int_0^{\infty}  \frac{2\pi v'_\perp dv'_\perp dv''_\parallel}{(\omega_{sn}-k_z v''_\parallel )} \Pi_{s\sigma32}=\sum_{n=-\infty}^{\infty} \Big\{  [ n\omega_{cs} \frac{ A_{nbs\sigma}}{v_{\perp ts\sigma}^2}\frac{Z_1}{k_z}+A_{n0\sigma} \frac{ v_{zts}Z_2 }{v_{zts}^2}]v_{dsy}+ [ n\omega_{cs} \frac{ A_{nbs\sigma}}{v_{\perp ts\sigma}^2}\frac{Z_0}{k_zv_{zts}}+A_{n0\sigma} \frac{Z_1 }{v_{zts}^2}]v_{dsy}v_{dsz} +i [(n\omega_{cs} -i\nu_s )\frac{ B_{nbs\sigma}}{v_{\perp ts\sigma}}\frac{Z_1}{k_z}+B_{n0\sigma} \frac{  v_{\perp ts\sigma}Z_2}{v_{zts} }]+i v_{dsz} [(n\omega_{cs} -i\nu_s )\frac{ B_{nbs\sigma}}{v_{\perp ts\sigma}}\frac{Z_0}{k_zv_{zts}}+B_{n0\sigma} \frac{ v_{\perp ts\sigma}Z_1}{v_{zts}^2}]   \Big\} $.
\end{itemize}
where the  $k_z$ in the numerator of $v_{dsy}$ term should be removed, i.e., $k_z v_{zts}Z_2$ should be $v_{zts}Z_2$. Otherwise, the dimension/unit is incorrect. The terms $v'_\perp\Pi_{s\sigma32}$ in page 360 and  $P_{s\sigma32}^m$ in page 364 should also be updated accordingly. This will affect the final matrix in the code for $v_{dsy}\neq0$ cases.

\end{document}